# Testing modified Newtonian dynamics with LISA Pathfinder

Christian Trenkel[1,*], Steve Kemble[1], Neil Bevis[2], and Joao Magueijo[2]
[1]Astrium Ltd, Gunnels Wood Road, Stevenage SG1 2AS, United Kingdom
[2]Theoretical Physics, Blackett Laboratory, Imperial College, London SW7 2BZ, United Kingdom

We suggest that LISA Pathfinder could be used to carry out a direct experimental test of MOND, in just a few years' time. No modifications of the spacecraft are required, nor any interference with its nominal mission. The basic concept is to fly LISA Pathfinder through the region around the Sun-Earth saddle point, in an extended mission phase, once the original mission goals are achieved. We examine various strategies to reach the saddle point, and find that the preferred strategy, yielding relatively short transfer times of just over one year, probably involves a lunar fly-by. LISA Pathfinder will be able to probe the intermediate MOND regime, ie the transition between deep MOND and Newtonian gravity. We present robust estimates of the anomalous gravity gradients that LISA Pathfinder should be exposed to, assuming the standard MOND interpolating function. The spacecraft speed and spatial scale of the MOND signal combine such that the signal falls precisely into LISA Pathfinder's measurement band. We find that if the gravity gradiometer on-board the spacecraft achieves its nominal sensitivity, these anomalous gradients could not just be detected, but measured in some detail. Conversely, given the large predicted signal, it is hard to see how MOND could be made compatible with a null result from such an experiment, while still accounting for galactic rotation curves.

## I. INTRODUCTION

The existence of Dark Matter was first postulated over 70 years ago, in order to reconcile observed galactic dynamics with standard gravitational theory [1]. Since then, Dark Matter has found its way into standard cosmology, and its abundance is thought to exceed that of ordinary baryonic matter by a factor of between five and six. However, despite decades of concerted effort, the direct search for Dark Matter particles remains essentially fruitless, and its nature is still one of the outstanding problems in cosmology.

In principle, the original problem with gravitational dynamics can also be solved by accepting deviations from standard gravitational theory. However, the complete vindication of General Relativity in all the experimental tests it has been subjected to so far, means that any alternative theory needs to be careful not to be inconsistent with those experimental tests, and will therefore almost by definition make predictions that deviate from General Relativity only in extreme environments (extreme in some characteristic parameters), and which are therefore not easily accessible to direct experimentation.

Modified Newtonian Dynamics (MOND) was originally proposed as a phenomenologically motivated scheme to explain galactic rotation curves without Dark Matter [2]. In essence, MOND postulates a deviation from Newtonian gravitation in systems (eg galaxies, or galaxy clusters) where the acceleration of the system falls below a certain threshold, of order $10^{-10}$ms$^{-2}$. This is indeed the case in the outskirts of many galaxies, and also on intergalactic scales. Much more recently, MOND was incorporated into a relativistic gravitational theory, the so-called Tensor-Vector-Scalar (TEVES) theory [3]. TEVES reproduces the MOND phenomenology in the non-relativistic limit.

At first sight, prospects of subjecting MOND to a direct test within our Solar System appear slim: at 1AU, the Sun's gravitational acceleration is around $6\times10^{-3}$ms$^{-2}$, and even Pluto is still attracted to the Sun at a rate of the order of $10^{-6}$ms$^{-2}$. Recently, however, it was pointed out that MOND leads to potentially measureable anomalous gravitational forces and force gradients in regions around gravitational saddle points within the Solar System, ie around points where the total gravitational acceleration of solar system bodies cancels [4]. In these localised regions an environment exists characterised by a very low total gravitational acceleration, similar (in this sense) to the outskirts of galaxies. In [4], reference is made to LISA Pathfinder (LPF), a technology demonstrator for LISA (Laser Interferometer Space Antenna) with a highly sensitive gravity gradiometer on-board, and it is suggested that an instrument such as that on LPF could in principle be used to detect anomalous MOND gravity gradients.

It has to be acknowledged that MOND (and TEVES) are far from being universally accepted. A dedicated space mission to test its predictions would therefore be unlikely to find sufficient support from the scientific community, due to the high cost. In this paper, our main aim is not to make a case for MOND. Our main aim is to take the thoughts above one step further, by suggesting that *LPF itself* could be used to conduct a direct test of MOND, without any changes to the hardware, or any interference with its nominal mission. LPF is a space mission already deep into its implementation phase, and has a planned launch date only a few years away.

Because of the existing reservations about MOND, we believe that LPF probably presents us with the only opportunity to carry out such a test in the foreseeable future, at a relatively modest cost (as far as space missions are concerned).

In order to convince ourselves that the above is a feasible proposition, we need to establish the following:

- LPF can be navigated through the vicinity of a gravitational saddle point once its nominal mission is completed
- The anomalous gravity gradients predicted by MOND can be detected and measured using the gravity gradiometer on board LPF

The remainder of the paper is organised as follows: sections II and III will address the first and second of the issues above, respectively, while section IV provides a summary and conclusions.

## II. FLYING LISA PATHFINDER THROUGH THE REGIONS AROUND GRAVITATIONAL SADDLE POINTS

### A. Gravitational Saddle Points in the Sun-Earth-Moon System

Within the Sun-Earth-Moon system, two distinct gravitational saddle points (SPs) exist, and therefore two potential targets for LPF. The first one is the Sun-Earth SP, which, ignoring perturbations due to the Moon and other bodies, is located on the Earth-Sun line, at a distance of approximately 259000km from Earth towards the Sun. The second SP is effectively the Moon-Sun SP, however the perturbation due to Earth is so large, that this is more a three-body SP rather than a two-body one. One can best visualise this SP as being nominally located on the Moon-Sun line, at a distance of approximately 30000km form the Moon towards the Sun. The perturbation of the Earth then results in a deflection of the SP away from this line, during the course of a lunar month, by up to 25° towards Earth. A comprehensive analysis of these two saddle points, and of the regions around them in which MOND effects may be measureable, can be found in [5].

In the interest of brevity, we present here just the main features of the two potential targets that will allow us to assess the pros and cons of targeting one or the other, and select one target for further investigation. Three figures of merit have been chosen:

(1) the size of the target region in which MOND effects will be measureable. This size is relevant for various reasons: Firstly, it will affect how easy it is to navigate LPF through it in the first place, and also how robust the signal prediction will be with respect to navigational errors. Secondly, the size of the region, combined with its own motion and the spacecraft speed, will determine the temporal variation of the signal.

(2) the number of opportunities, or time slots ("launch windows"), during which LPF can leave its nominal orbit to reach the target. Clearly, the larger the number of possible departure times, the higher the flexibility of the mission. However, this should be considered together with the transfer time to the target.

(3) the typical speed of the SPs in a coordinate system in which Earth and Sun remain fixed. Not only the lunar SP, which approximately travels with the Moon around its orbit, but also, to less extent, the Sun-Earth SP are dynamic SPs, and therefore their speed needs to be taken into account to predict the temporal signal variation as mentioned above.

The following table shows a comparison between the two targets with respect to these figures of merit (for additional details please refer to [5]):

|  | **Sun-Earth SP** | **Lunar SP** |
|---|---|---|
| **Characteristic Target Region Size** | ≈380km (for most of the time, down to about 330km for short periods every month) | ≈50km (varies between 30km and 80km with lunar phase) |
| **"Launch Windows" from nominal LPF orbit** | Many: SP motion fairly limited and therefore more "stable" target | Fewer: departure time needs to be synchronised with lunar motion |
| **Speed relative to Earth – Sun system** | <<1km/s for most of the time, exceeding 0.1km/s only for short periods every month | Typically of order 1km/s as it follows the Moon on its orbit around Earth |

Table I: Main comparative features of Sun-Earth SP and Lunar SP

For the work presented here, we decided to focus on the Sun-Earth SP because it presents a more predictable, stable and larger target, with, potentially increased flexibility as far as departure times are concerned. In future work, however, it may be worth exploring the lunar SP further.

### B. Assumptions about LISA Pathfinder

The search for trajectories that will take the spacecraft from its nominal orbit around L1 through the target region identified above has to take into account various constraints, derived from the requirement to use LPF as built, and not to interfere with the nominal mission. The main constraints are:

- The nominal orbit around L1 should be used as starting point for any manoeuvres of the LPF spacecraft. Although this orbit is not completely pre-determined, and depends on the precise orbit insertion and subsequent evolution, it is sufficiently well known for our purpose.
- The possible deltaV manoeuvres are severely limited by the micropropulsion system on-board LPF– in particular in terms of thrust capability and propellant quantity.

- Lifetime: the instrument has been designed with a certain, finite lifetime in mind, limited for example by the vacuum level inside the gravity gradiometer sensors. For this reason, trajectories will not be propagated for more than 2 years.

The properties of the LPF micropropulsion system, comprising three clusters of four Field-Electric-Effect-Propulsion (FEEP) thrusters each, are particularly restrictive, and will therefore be discussed in some more detail. FEEPs are low-thrust, high precision thrusters, ideal to provide the drag-free environment required for LPF. Under nominal conditions, the thrusters provide 90μN each, and are arranged in clusters around the spacecraft to provide full 6 Degree-Of-Freedom attitude control. Taking into account a spacecraft mass of order 500kg, it is clear that maximum spacecraft accelerations in any particular direction are very low. Current analysis shows that in the direction of interest (see next section), a total spacecraft velocity change of 1m/s can be achieved in 21 days with all thrusters functional, or 33days assuming a single thruster failure. Given the slow acceleration, any significant trajectory correction manoeuvres need to be planned and executed well in advance, and will be most effective when applied at specific points along the trajectory such as apogees, when the spacecraft velocity is at a minimum; no last-minute course correction can be applied.

LPF propellant budgets indicate that following the nominal mission, the residual propellant should be enough to apply a total velocity change of between 6m/s and 15m/s. From the above consideration, it can be concluded that carefully timing and executing correction manoeuvres will impose more of a constraint on the trajectories that can be followed, than the total amount of available propellant.

C. Orbital Analysis: Trajectories available to LISA Pathfinder

The nominal LPF orbit is a large amplitude Lissajous orbit around L1 (in fact close to Halo orbit dimensions). Like all such orbits, it is unstable if adversely perturbed from its ideal state. This instability can be exploited to enable the spacecraft to leave the orbit with only a small manoeuvre. A velocity change applied in a critical direction of 28 deg from the Earth-Sun direction ensures that the spacecraft reaches an 'unstable manifold'. This means that over a period comparable with that of the Lissajous orbit, it will strongly diverge from its nominal motion. Possible motions include low energy escape trajectories and also trajectories returning to Earth with a low perigee.

The problem to be solved is to find an appropriate manoeuvre after which it is possible to target the Sun-Earth gravitational equilibrium zone. Ignoring in the first instance the effect of the Moon, the SP is located on the Earth-Sun line, at a distance from Earth (in Earth to Sun direction) of 259000km.

It is important to emphasise that, due to the chaotic nature of the problem, and its susceptibility to small perturbations, it is impossible at this stage to predict the optimal trajectory that LPF should follow. This is because some parameters which affect the trajectory, such as the exact starting orbit, or the exact solar radiation pressure on the spacecraft, can only be determined with sufficient accuracy once LPF is in its nominal orbit. Therefore, the final trajectory planning and optimisation can only be carried out then. The primary goal here is, instead, to demonstrate the existence of a family of solutions to the problem, and to gain some insight into the general trajectory properties.

The merit of any individual solution that is found can be assessed based on mainly two parameters: the SP "miss distance", ie the distance by which the actual SP is missed, and the transfer time, ie the time it takes from the initial departure from L1 to cross the SP target region.

A velocity change of 1m/s, as discussed above, can be achieved within 33 days at most. This will be the maximum deltaV that will be considered, and is possible in the slowly moving Lissajous orbit (in practice, the real LPF trajectory will be followed by continuous and iterative navigation and small trajectory correction manoeuvres, rather than just a single initial manoeuvre).

To indicate some typical trajectories available to LPF, and using the time of departure from the L1 orbit as single search parameter, Figure 1 shows a range of possible trajectories resulting from 1m/s deltaV manoeuvres.

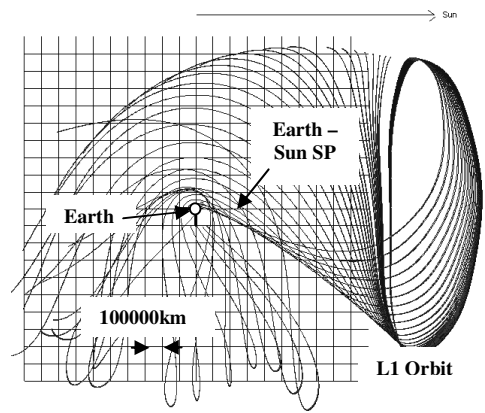

Figure 1: A range of orbits leaving the nominal Lissajous orbit, with a deltaV of 1m/s. The time application for each trajectory is incremented in 5 day steps. The grid is in the ecliptic, using an Earth-Sun rotating reference.

The large out-of-ecliptic motion of the nominal orbit makes it difficult to meet approach the target region around the SP on the first 'leg' of the trajectory, where the spacecraft starts its return towards Earth. However the subsequent motion after Earth perigee is weakly bound and offers numerous further possibilities to reach the target region.

Additional deltaVs can also be applied in this manoeuvre phase. However perigee manoeuvres are relatively ineffective due to the very low thrust/mass ratio, therefore manoeuvres close to apogee can be effectively used to modify the motion.

As will be seen, an interesting option is to target a Lunar fly-by, as this offers a further effective manoeuvre – less constrained by the micropropulsion system properties – that helps to target the region of interest.

Expanding the search method to two parameters, it is possible to locate trajectories that pass close to the target region. The control parameters employed are the time at which the deltaV is applied, as before, and the size of the deltaV. Some results obtained with a coarse two parameter search results are shown in Figure 2:

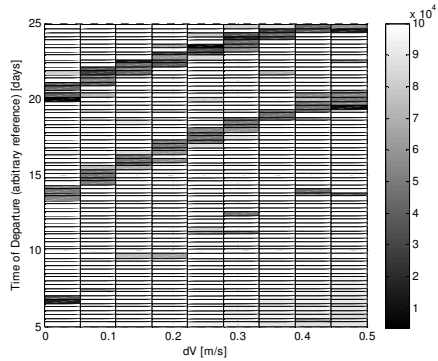

Figure 2: Sun-Earth SP miss distance (in km) obtained using a coarse two parameter search, varying time of departure and magnitude of deltaV manoeuvre.

As can be seen, the search indicates some clustering of low miss distances with time at which the manoeuvre is made, relative to an arbitrary reference. Numerous local minima are found, which could in principle be investigated further. For information, the lowest miss distances amongst the solutions found above are of order 3000km.

In general, we observe that relatively fast transfer times (< 1 year) only achieve large miss distances of 20000km to 30000km, while smaller distances, below 5000km, are achieved only for significantly longer transfer times (of the order of 1.5 years). Smaller miss distances could be achieved with additional manoeuvres, but the main goal here is to understand the nature of the problem, not to find a particular solution.

However, trajectories that include Lunar fly-bys do not follow this general trend, and we believe that these present the best opportunity to combine low miss distances with reasonably short transfer times. Although fewer launch windows may exist depending on the position of the Moon, this is probably offset by the shorter transfer duration.

The solution with the lowest miss distance found assuming a single deltaV manoeuvre is shown in Figure 3, mainly for illustration purposes. Again, it needs to be stressed that the actual LPF trajectory would be very different. In this particular case, the miss distance that is achieved is around 600km, after a transfer time of 410 days (< 14 months).

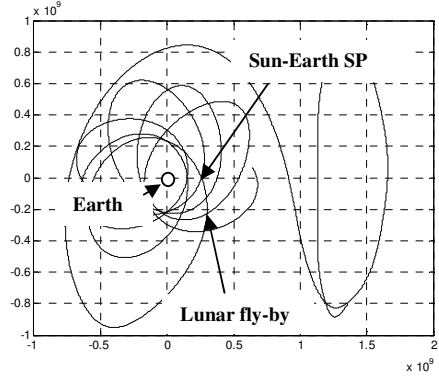

Figure 3: Trajectory including lunar fly-by after 300 days, with miss distance of 600km after total transfer time of 410days.

Numerous options exist to achieve closer approaches. The principle involves the application of further control parameters, which in general means a second manoeuvre. This could be applied at some time after the first manoeuvre, whilst still close to the initial Lissajous orbit, or at one of the subsequent apogees.

The miss distance will ultimately be limited by navigation errors, ie a combination of the uncertainty in the spacecraft position itself, and in the applied deltaV manoeuvres. Conservatively, it can be stated that a miss distance of 10-20km will be achievable.

D. Additional considerations and conclusions

In the previous sections we have shown that LPF *can* in all likelihood be made to cross the region around the Sun-Earth SP, following its nominal mission. The nominal LPF orbit is relatively unstable, which means that only very small deltaV manoeuvres, compatible with the micropropulsion system on-board, are required to achieve this. The final optimised trajectory cannot be determined from ground, and will have to be planned once LPF is on-station. The achievable SP miss distance will be limited, ultimately, by navigational inaccuracies, and is estimated, conservatively, to be between 10 and 20km. We have also found that exploiting a lunar fly-by to execute an additional manoeuvre helps to reduce the transfer times from L1 orbit to the target region to about 14 months. This duration is considered to be compatible with the anticipated LPF lifetime.

An additional quantity of interest, as will be seen, is the spacecraft speed when crossing the target region. Given our limited propulsion capabilities, the spacecraft is effectively coasting, and its speed simply given by the exchange of kinetic and potential gravitational energy during the trajectory. Although small variations are possible, the spacecraft speed can be predicted to be of the order of 1.5km/s.

One final comment to be made is that all trajectories found so far offer only a single pass through the target region, at least within the 2 years over which they were propagated. Any attempt to detect anomalous MOND signals with LPF will be a one-off experiment.

### III. SENSITIVITY OF LISA PATHFINDER TO MOND ANOMALOUS GRAVITY GRADIENTS

#### A. Anomalous Signal Calculation

In order to assess whether LPF will be able to detect the anomalous gravity gradients, we need to be aware of which signals LPF will be able to measure. The gravity gradiometer on-board LPF consists of two freely floating test masses, separated by a baseline of just under 40cm. The high sensitivity interferometric readout is only available along the axis joining the two test masses. Capacitive readout exists for the other translational degrees of freedom but this is much less sensitive. This means that LPF will be capable of measuring, with high precision, just one inline gradient. In addition, for thermal stability reasons, the spacecraft attitude needs to be such that its solar array is always approximately perpendicular to the Sun. Given the accommodation of the gradiometer on-board the spacecraft, this means that the gravity gradient measured by LPF has to be in a plane perpendicular to the Sun vector (see [6] for a detailed overview of the spacecraft and its payload). It is therefore MOND gradients in this plane (for symmetry reasons, they are expected to be very similar) that we need to estimate.

A robust calculation of these gradients is obviously required. Two regimes are discussed in [4], in which analytical predictions can be made relatively easily: firstly the "deep MOND" regime, characterised by external gravitational accelerations less than $10^{-10}$ms$^{-2}$, and secondly the regime where deviations from Newtonian gravity can be treated as small perturbations. The intermediate regime in between these two is called "strong MOND" regime. The deep MOND regime is given only in a small region of negligible size around the SP, and inaccessible to LPF. The perturbative regime where MOND effects are small starts at a distance from the SP described earlier as characteristic target region size, of approximately 380km for the Sun-Earth SP [4,5].

Newtonian gradients around the Sun-Earth SP are around $4 \times 10^{-11}$s$^{-2}$ in the Sun-Earth direction. Assuming that LPF can be made to miss the SP by 10km, and assuming best-case Newtonian gradients of no more than $2 \times 10^{-11}$s$^{-2}$, the range of accelerations that LPF will be exposed to, while inside this target region, will vary between about $2 \times 10^{-7}$ms$^{-2}$ at closest approach, and $7.6 \times 10^{-6}$ms$^{-2}$ at the boundary. It can be concluded that during the entire crossing of the target region, LPF will be exposed to MOND effects which do not fall into either regime, and can therefore not be estimated analytically. Furthermore, since LPF will be exploring this intermediate, strong MOND regime, the choice of interpolating function becomes fundamental. In [5], a robust numerical method has been developed to deal with precisely this intermediate regime: it calculates MOND effects on a three-dimensional grid, and is capable of including an arbitrary number of gravitational sources outside the grid. All signal estimates presented here have been generated using this method, and the "standard" interpolating MOND function has been assumed [3,5]

Since the exact LPF trajectory is not yet known, a typical velocity vector has been assumed for the spacecraft through the target region, with an "elevation" (relative to the ecliptic) of 45°, and an "azimuth" (relative to the Sun-Earth line) of 45°.

In order to further assess the robustness of the signal prediction, the signal has been calculated when the following parameters were varied:

- the SP miss distance (between 0 and 400km, just outside the target region)
- lunar position, exploiting the generality of the method developed in [5]

The numerical method calculates anomalous fields and field gradients on a spatial grid; with the assumed spacecraft trajectory through this grid and its speed, these spatial variations are converted into temporal variations. To be conservative, a spacecraft speed of 1km/s has been assumed.

It should be noted that for numerical reasons – the finite distance between discrete grid points –, the "0km" miss distance displayed below was in fact closer to 1 or 2km, but it is not thought that this introduces significant errors.

Figure 4 shows the calculated temporal MOND gradients that LPF would see, for several miss distances up to 400km. These results correspond to the new Moon case; the signal dependence on lunar phase has been investigated and found to be small [5].

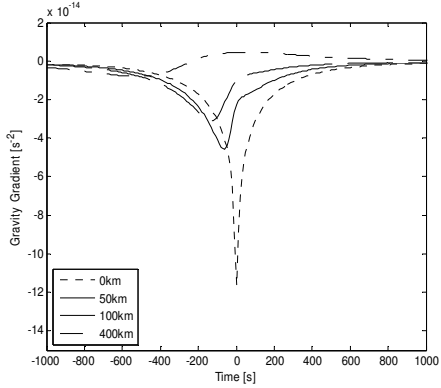

Figure 4: Anomalous MOND gradients for various SP miss distances. The origin on the time axis corresponds to the closest approach.

The above plot also serves as an indication of the signal sensitivity with respect to navigational uncertainties. In what follows, we will assume the signal strength for the 50km miss distance, as a worst case signal. With LPF targeting the SP itself, but perhaps missing it by 10-20km, the signal is certain to be larger than that estimated for the 50km case.

### B. LISA Pathfinder Gradiometer Sensitivity and Additional Noise Sources

Two main noise sources have been identified that could potentially limit the ability of LPF to detect the MOND signals calculated above:

- Intrinsic instrument noise
- External Newtonian background

For the intrinsic gradiometer sensitivity, we will assume a nominal noise spectral density of $1.5 \times 10^{-14} s^{-2}/\sqrt{Hz}$ as quoted in [7], and will further assume that this applies between 1 and 10mHz. Below 1mHz we will assume a $1/f$ increase in the amplitude spectral density, and above 10mHz we will assume an increase in proportion to $f^2$.

External Newtonian gradients around the saddle point are around $2-4 \times 10^{-11} s^{-2}$, and therefore at least two orders of magnitude larger that the predicted MOND gradients. For two reasons this is not a problem, as we shall see: Firstly, only the *uncertainty* in the Newtonian background is really a noise source: knowledge of the spacecraft trajectory with a typical uncertainty of 10km is equivalent to unknown gradients of the order of $10^{-15} s^{-2}$. Secondly, the temporal variation of the Newtonian gradients is very different from the MOND ones, and a simple polynomial fit can be used to remove it from the data.

### C. Overall Signal to Noise Comparison

Having obtained a robust signal prediction and considered the main noise sources, we can now compare the two. This can be done both in the time and frequency domains, and we believe both approaches provide useful insights.

Starting in the time domain, we have generated characteristic time series of the total gravitational gradients that LPF would see, both in the presence and absence of MOND. In the absence of MOND, the total signal would be given by the sum of the Newtonian background and instrument noise, and in its presence we simply add the MOND prediction as presented above. We have simulated the instrument noise by generating a time series from the noise spectral density described above between 50μHz and 50mHz, with random Fourier phases.

Figure 5(a) shows the raw data, figure 5(b) shows the same data after being low-pass filtered using a simple 10-point moving average. Clearly the MOND signal can be recovered from the data very well.

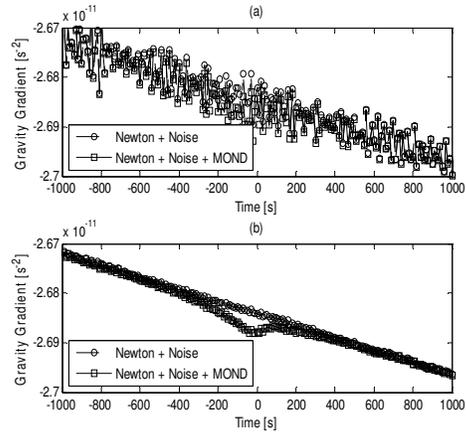

Figure 5 (a) Raw LPF gravity gradient in presence and absence of MOND, (b) same data, low-pass filtered using simple 10-point moving average.

From figure 5, the very different temporal variations of Newtonian and MOND gradients, and also instrument noise, become apparent: MOND gradients are significant on a timescale of the order of 500-1000s, while the Newtonian gradients show variations on much longer timescales. The instrument noise is seen to dominate at high frequencies. There is, then, a frequency window near 1mHz within which the predicted MOND signal dominates.

The frequency domain approach is perhaps less indicated for the analysis of a single event such as the SP region crossing, but helps to further clarify the frequencies over which the various contributions dominate. Figure 6 shows the amplitude spectral density of the MOND signals shown in figure 4, compared to the nominal LPF instrument noise spectral density.

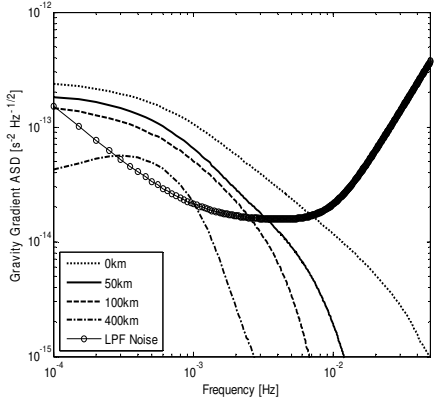

Figure 6: Amplitude spectral densities of calculated MOND signals for various miss distances, compared to the nominal intrinsic LPF gradiometer sensitivity.

As can be seen, the calculated MOND variations conveniently overlap with the frequency band over which LPF should achieve its highest sensitivity. This is a fortunate combination of the spacecraft speed and the scale over which MOND effects become significant. In fact, a more realistic spacecraft speed of 1.5km/s (1km/s was assumed above) would further shift the signal into the nominal LPF measurement band.

As is evident from Figure 6, the MOND gradients should not just be detectable with LPF, but *measureable with considerable signal to noise ratio*. For a pessimistic miss distance of 50km and a conservative spacecraft speed of 1km/s, the power in the signal exceeds the power in the noise by at least a factor of 10 in a narrow frequency band around 1mHz.

It is concluded that if LPF achieves its nominal performance, then it can be used to subject the MOND predictions obtained assuming the standard interpolating function to a very stringent test.

### IV. SUMMARY AND CONCLUSIONS

In this paper, we have investigated whether LISA Pathfinder, a space mission with a highly sensitive gravity gradiometer on-board, and with a launch date only a few years away, could be used to carry out a direct experimental test of MOND.

We have shown that in principle, following its nominal mission, LPF can be made to fly through the region around the Sun-Earth saddle point where the MOND effects become measureable. Although LPF cannot be used to explore the deep MOND regime, it can certainly test the transition regime between MOND and Newtonian Gravity. A powerful numerical technique has been developed which has allowed us to make robust estimates of the anomalous gradients that LPF would be exposed to near the saddle point, based on the standard interpolating function of MOND. Assuming that the LPF gradiometer achieves its nominal performance, we have shown that the sensitivity is adequate to not just detect, but to measure the predicted MOND gradients in detail. Given the strength of the MOND signal, should LPF detect absolutely no anomalies during its crossing of the saddle point region, it is hard to see how MOND could be compatible with such a null result, and still be used to explain galactic rotation curves. Of course, it almost goes without saying that a positive experimental verification of MOND would represent a major breakthrough in physics and cosmology.

Since the idea presented here is based on an existing mission, and the proposed test would therefore be relatively low cost – certainly compared to, say, a dedicated mission –, we believe that it is worth exploring further.

### ACKNOWLEDGEMENTS

We would like to acknowledge useful discussions with Tim Sumner and all members of the LISA Pathfinder Science Working Team. C.T and S.K. would like to acknowledge financial support from Astrium Ltd.